\documentclass[twocolumn]{aastex6}
\usepackage{bm}
\usepackage{amsmath}

\usepackage[section]{algorithm}
\usepackage{algpseudocode}

\DeclareMathOperator*{\argmin}{argmin}
\newcommand{\V}{\bm}
\accepted{January 24 2017}
\slugcomment{Accepted for publication in the Astronomical Journal}
\shorttitle{Super-resolution Full Polarimetric Imaging with Sparse Modeling}
\shortauthors{K.~Akiyama et al.}
\newcommand{\haystack}{1}
\newcommand{\naoj}{2}
\newcommand{\ism}{3}
\newcommand{\sokendaiism}{4}
\newcommand{\smith}{5}
\newcommand{\utokyo}{6}
\newcommand{\perimeter}{7}
\newcommand{\uwaterloo}{8}
\newcommand{\mpe}{9}
\newcommand{\radboud}{10}
\newcommand{\sokendainaoj}{11}
\newcommand{\cfa}{12}
\newcommand{\jsps}{13}
\begin{document}
\title{Super-resolution Full Polarimetric Imaging for Radio Interferometry\\with Sparse Modeling}
%
\author{Kazunori Akiyama\altaffilmark{\haystack,\naoj,\jsps}}
\author{Shiro Ikeda\altaffilmark{\ism,\sokendaiism}}
\author{Mollie Pleau\altaffilmark{\haystack,\smith}}
\author{Vincent L. Fish\altaffilmark{\haystack}}
\author{Fumie Tazaki\altaffilmark{\naoj}}
\author{Kazuki Kuramochi\altaffilmark{\utokyo,\naoj}}
\author{Avery Broderick\altaffilmark{\perimeter,\uwaterloo}}
\author{Jason Dexter\altaffilmark{\mpe}}
\author{Monika Mo{\'s}cibrodzka\altaffilmark{\radboud}}
\author{Michael Gowanlock\altaffilmark{\haystack}}
\author{Mareki Honma\altaffilmark{\naoj,\sokendainaoj}}
\author{Sheperd S. Doeleman\altaffilmark{\cfa}}
%
\altaffiltext{\haystack}{Massachusetts Institute of Technology, Haystack Observatory, 99 Millstone Road, Westford, MA 01886, USA}
\altaffiltext{\naoj}{Mizusawa VLBI Observatory, National Astronomical Observatory of Japan, 2-21-1 Osawa, Mitaka, Tokyo 181-8588, Japan}
\altaffiltext{\ism}{Department of Statistical Science, School of Multidisciplinary Sciences, Graduate University for Advanced Studies,  10-3 Midori-cho, Tachikawa, Tokyo 190-8562, Japan}
\altaffiltext{\sokendaiism}{Graduate University for Advanced Studies, 10-3 Midori-cho, Tachikawa, Tokyo 190-8562, Japan}
\altaffiltext{\smith}{Department of Astronomy, Smith College, Northampton, MA 01063, USA}
\altaffiltext{\utokyo}{Department of Astronomy, Graduate School of Science, The University of Tokyo, 7-3-1 Hongo, Bunkyo-ku, Tokyo 113-0033, Japan}
\altaffiltext{\perimeter}{Perimeter Institute for Theoretical Physics, 31 Caroline Street, North Waterloo, Ontario N2L 2Y5, Canada}
\altaffiltext{\uwaterloo}{Department of Physics and Astronomy, University of Waterloo, 200 University Avenue West, Waterloo, Ontario N2l 3G1, Canada}
\altaffiltext{\mpe}{Max Planck Institute for Extraterrestrial Physics, Giessenbachstr.\ 1, 85748 Garching, Germany}
\altaffiltext{\radboud}{Department of Astrophysics/IMAPP, Radboud University Nijmegen, P.O. Box 9010, 6500 GL Nijmegen, The Netherlands}
\altaffiltext{\sokendainaoj}{Department of Astronomical Science, School of Physical Sciences, Graduate University for Advanced Studies, 2-21-1 Osawa, Mitaka, Tokyo 181-8588, Japan}
\altaffiltext{\cfa}{Harvard-Smithsonian Center for Astrophysics, 60 Garden Street, Cambridge, MA 02138, USA}
\altaffiltext{\jsps}{\url{kazu@haystack.mit.edu}; JSPS Postdoctoral Fellow for Research Abroad}
%
\begin{abstract}
We propose a new technique for radio interferometry to obtain super-resolution full polarization images in all four Stokes parameters using sparse modeling. The proposed technique reconstructs the image in each Stokes parameter from the corresponding full-complex Stokes visibilities by utilizing two regularization functions: the $\ell _1$-norm and total variation (TV) of the brightness distribution. As an application of this technique, we present simulated linear polarization observations of two physically motivated models of M87 with the Event Horizon Telescope (EHT). We confirm that $\ell _1$+TV regularization can achieve an optimal resolution of $\sim 25-30$\% of the diffraction limit $\lambda/D_{\rm max}$, which is the nominal spatial resolution of a radio interferometer for both the total intensity (i.e. Stokes $I$) and linear polarizations (i.e. Stokes $Q$ and $U$). This optimal resolution is better than that obtained from the widely used Cotton-Schwab CLEAN algorithm or from using $\ell _1$ or TV regularizations alone. Furthermore, we find that $\ell _1$+TV regularization can achieve much better image fidelity in linear polarization than other techniques over a wide range of spatial scales, not only in the super-resolution regime, but also on scales larger than the diffraction limit. Our results clearly demonstrate that sparse reconstruction is a useful choice for high-fidelity full-polarimetric interferometric imaging.
\end{abstract}

\section{Introduction}\label{sec:1}
Polarization is a unique tracer of the magnetic field and magnetized plasma distribution in the universe. The distribution of magnetic field lines can be inferred from linear polarization in a variety of sources, including synchrotron emission from non-thermal or relativistic thermal electrons in high-energy objects \citep[e.g.][]{pacholczyk1970}, maser emission from star-forming regions or evolved stars \citep[e.g.][]{fish2006}, and thermal emission partially absorbed by aligned dust grains \citep[e.g.][]{girart2006,girart2009}. Polarized emission also contains information about the magnetized plasma along the line of sight \citep[e.g. Faraday Tomography;][]{burn1966,brentjens2005} via Faraday rotation of linear polarization or Faraday conversion from linear to circular polarization \citep{legg1968,pacholczyk1970,jones1977}. Recent theoretical and observational studies suggest that linear polarization can be a unique tracer of the dust evolution in proto-planetary disks \citep{kataoka2015,kataoka2016}.

High-resolution imaging of polarized emission is therefore a fundamental part of the modern observational toolkit in astronomy. The angular resolution of a telescope (often referred to as ``beam size'' in radio astronomy and ``diffraction limit'' in optical astronomy) is given by $\theta \approx \lambda/D$, where $\lambda$ and $D$ are the observing wavelength and the diameter of the telescope, respectively. At radio wavelengths, interferometry is the most effective approach to obtain high angular resolution.  The nominal resolution of an interferometer is given by $\theta \approx \lambda/D_{\rm max}$, where $D_{\rm max}$ is the maximum length of the baseline between two telescopes, projected in the plane normal to the direction of observation.  Of all observational techniques across the electromagnetic spectrum, radio interferometry provides the imaging capability at the finest angular resolution \citep[e.g.][]{thompson2001}. In particular, very long baseline interferometry (VLBI), which utilizes intercontinental baselines (or even baselines to space), has achieved the highest angular resolution in the history of astronomy. 

The Event Horizon Telescope \citep[EHT;][]{doeleman2009b} is a ground-based VLBI array with an angular resolution of a few tens of microarcseconds at short/sub-millimeter wavelengths ($\lambda \lesssim 1.3$~mm, $\nu \gtrsim 230$~GHz) \citep[e.g.][]{doeleman2008,doeleman2012,fish2011,fish2016,lu2012,lu2013,akiyama2015,johnson2015}. The EHT resolves compact structures of a few Schwarzschild radii ($R_s$) in the vicinity of the supermassive black holes in the Galactic Center source Sgr A* \citep{doeleman2008,fish2011,fish2016,johnson2015} and the nucleus of M87 \citep{doeleman2012,akiyama2015}. Direct imaging of these scales will be accessible in the next few years with technical developments and the addition of new (sub)millimeter telescopes such as the Atacama Large Submillimeter/millimeter Array (ALMA) to the EHT \citep[e.g.][]{fish2013}. Polarimetric imaging with the EHT will be especially transformational, opening a new field to study the properties of the magenetic field distribution and magnetized plasma in the regions of strong gravitation. Indeed, early EHT observations of Sgr A* resolves ordered and time-variable magnetic fields on $R_s$ scales \citep{johnson2015}. High-fidelity images of the linearly polarized emission will be crucial for understanding processes of black hole accretion and jet formation.

The imaging problem of interferometry is formulated as an underdetermined linear problem (see \S\ref{sec:2.1}) of reconstructing an image from complex visibilities that represent Fourier components of the source image. The CLEAN algorithm \citep{hogbom1974} and its variants \citep[e.g.][]{clark1980,schwab1984} have been the most successful and widely used algorithms in radio interferometry. CLEAN was independently rediscovered as the Matching Pursuit algorithm \citep[MP;][]{mallat1993} and has been widely used in many other fields to derive a sparse solution ${\bf x}$ of an underdetermined linear problem ${\bf y}={\bf Ax}$, where ${\bf y}$ and ${\bf A}$ are observational data and observation matrix, respectively. For real data with noise, this can be mathematically described by
\begin{equation}
\min _{\bf x} ||{\bf y-Ax}||^2_{2}\,\,\,{\rm subject\,\,to\,\,}||{\bf x}||_{0} \leq T, \label{eq:sparse}
\end{equation}
where $||{\bf x}||_{p}$ is the $\ell _p$-norm of the vector ${\bf x}$ given by
\begin{equation}
||{\bf x}||_{p} = \left( \sum_i |x_i|^p \right) ^{\frac{1}{p}}
\end{equation}
for $p>0$, and defined as the number of non-zero components for $p=0$. The term to be minimized is the traditional $\chi ^2$ term, and $T$ is a threshold for the $\ell _0$-norm representing the sparsity of the solution. Thus, the solution is equivalent to minimizing the $\chi ^2$ term within a given sparsity.
A direct approach to solve this equation is to try all possible combinations of zero components of ${\bf x}$ one-by-one. However, the computational cost of this exhaustive search is so large that
it is intractable for large dimensional $\V{x}$.  CLEAN, MP and their variants are select a non-zero element one-by-one and incrementally in a greedy manner in order to solve this problem efficiently.

A popular relaxation of sparse reconstruction comes from replacing the $\ell _0$-norm with the $\ell _1$-norm as
\begin{equation}
\min _{\bf x} ||{\bf y-Ax}||^2_{2}\,\,\,{\rm subject\,\,to\,\,}||{\bf x}||_{1} \leq T,
\end{equation}
which is known as LASSO \citep[least absolute shrinkage and selection operator;][]{tibshirani1996}. This is a convex relaxation of Eq.~(\ref{eq:sparse}), and can be transformed in the Lagrange form,
\begin{equation}
\min_{\bf x} \left( ||{\bf y} - {\bf A} {\bf x}||_{2}^2 + \Lambda _{\ell} ||{\bf x}||_{1} \right).
\end{equation}
Many efficient algorithms have been proposed to solve LASSO \citep[e.g., the fast iterative shrinkage-thresholding algorithm (FISTA);][]{beck2009b}. This method has become popular in many fields such as medical imaging, particularly after the appearance of compressed sensing \citep[also known as compressive sensing;][]{donoho2006,candes2006} techniques, which have shown that LASSO can solve many ill-posed linear problems accurately if the solution vector is sparse --- the number of elements with nonzero value is much small compared to its dimension. We have applied LASSO to Stokes $I$ imaging with radio interferometry \citep{honma2014,ikeda2016,akiyama2016} and found that LASSO can potentially reconstruct structure on scales $\sim 4$ times finer than $\lambda/D_{\rm max}$ \citep{honma2014}. Techniques of compressed sensing are beginning to be used in other fields of radio interferometry \citep[see][and references therein]{garsden2015}.

A critical assumption in techniques with $\ell _1$ regularization is that the solution (i.e., the true image) is sparse. If the number of pixels with nonzero brightness is not small compared to the number of data points, simple $\ell _1$-norm regularization may reconstruct an image that is too sparse. This situation can arise when reconstructing an extended source or even for a compact source if the imaging pixel size is set to be much smaller than the size of the emission structure. A promising approach to overcome this issue is to change the basis of the image to a more sparse one. Pioneering work in this area has made use of transforms to wavelet or curvelet bases, in which the image can be represented sparsely \citep[e.g.][]{li2011,carrillo2014,garsden2015,dabbech2015}. We have taken another approach by adding total variation (TV) regularization \citep[e.g.][]{wiaux2010,mcewen2011,uemura2015, chael2016}, which produces an image that is sparse in its gradient domain. TV regularization has been shown to be effective for imaging with visibility amplitudes and closure phases \citep[e.g.][]{akiyama2016} in the super-resolution regime finer than the diffraction limit.

In interferometric imaging, another class of widely-used imaging techniques is the Maximum Entropy Methods (MEM), utilizing different functions (named as ``entropy terms'') to regularize images \citep[see][for a review]{narayan1986}. Image reconstruction with MEM has been practically extended to polarimetry (\citealt{holdway1990,sault1999}; and see \citealt{chael2016} and \citealt{cohghlan2016} for a review of polarimetric MEM techniques).

Sparse modeling techniques utilizing $\ell _1$ and TV terms have heretofore been applied only to Stokes $I$ image reconstruction. In this paper, we extend the framework of sparse imaging techniques for radio interferometry with $\ell _1$ and TV regularizations to full-polarization imaging for the first time. As an example, we apply our new technique to simulated EHT data of the accretion and jet launching region immediately around the black hole of M87.

\section{The proposed method}\label{sec:2}
\subsection{Polarimetric imaging with radio intererometry}\label{sec:2.1}
The intensity distribution of the emission from the sky can be described with four Stokes parameters, $I$, $Q$, $U$ and $V$, which are all real. Stokes $I$ represents the total intensity of the emission, which is generally non-negative for astronomical images. On the other hand, $Q$ and $U$, which represent linear polarization, and $V$, which represents circular polarization, can take on negative values. Stokes $Q$ and $U$ are often combined into the complex quantity $P\equiv Q+iU$, where $|P|$ and $\chi=\arg(P)/2$ are the linear polarization intensity and the electric vector polarization angle (EVPA), respectively. 

A radio interferometer samples Fourier components of each Stokes parameter, known as the Stokes visibilities $\tilde{I}$, $\tilde{Q}$, $\tilde{U}$ and $\tilde{V}$ defined by
\begin{equation}
\tilde{S}(u,v)= \int dx dy\, S(x,y) \exp(-i2\pi (ux+vy)),\\
\end{equation}
where $S$ and $\tilde{S}$ represent a Stokes parameter and corresponding Stokes visibility (i.e. $S=I,Q,U,V$). Here, the spatial frequency $(u,v)$ corresponds to the baseline vector (in units of the observing wavelength $\lambda$) between two antennas (or receivers) projected to the tangent plane of the celestial sphere at the phase-tracking center.

Observed visibilities are discrete quantities, and the sky image can be approximated by a pixellated version where the pixel size is much smaller than the nominal resolution of the interferometer. The Stokes parameters can therefore be represented as discrete vectors ${\bf I}$, ${\bf Q}$, ${\bf U}$ and ${\bf V}$, related to the observed Stokes visibilities $\tilde{{\bf I}}$, $\tilde{{\bf Q}}$, $\tilde{{\bf U}}$ and $\tilde{{\bf V}}$ by a discrete Fourier transform ${\bf F}$:
\begin{equation}
\tilde{{\bf S}}= {\bf F} {\bf S}~~({\rm for}~{\bf S}={\bf I},~{\bf Q},~{\bf U},~{\bf V}). \label{eq:obseq}
\end{equation}
The sampling of Stokes visibilities is almost always incomplete. Since the number of visibility samples $\tilde{{\bf S}}$ is smaller than the number of pixels in the image, solving the above equation for the image ${\bf S}$ is an ill-posed problem. One or more regularizers must therefore be added to find a unique solution to equation (\ref{eq:obseq}).

\subsection{The Proposed Methods}\label{sec:2.2}

A natural extension of our previous work \citep{honma2014,ikeda2016, akiyama2016} to full polarimetric imaging is given by
\begin{equation}
{\bf S} = {\rm argmin}_{\bf S} \left( ||\tilde{{\bf S}} - {\bf F} {\bf S}||_{2}^2 + \Lambda _{\ell} ||{\bf S}||_{1} + \Lambda _{t} ||{\bf S}||_{\rm tv} \right) \label{eq:opteq}
\end{equation}
for each Stokes parameter (i.e. ${\bf S}={\bf I},~{\bf Q},~{\bf U},~{\bf V}$) and corresponding Stokes visibility (i.e.  $\tilde{{\bf S}}=\tilde{{\bf I}},~\tilde{{\bf Q}},~\tilde{{\bf U}},~\tilde{{\bf V}})$. This equation consists of the traditional $\chi ^2$ term, which represents deviations between the model image and observed visibilities, and two terms consisting of a regularizer and a regularization parameter.


One of the additional terms is $\ell _1$-regularization \citep[e.g.][]{honma2014}. $\Lambda _{\ell}$ is its regularization parameter, adjusting the degree of sparsity by changing the weight of the $\ell _1$-norm penalty. In general, a large $\Lambda_{\ell}$ prefers a solution with very few nonzero components, while a small $\Lambda_{\ell}$ imposes less sparsity. In this paper, we use the normalized regularization parameter $\tilde{\Lambda}_{\ell}$
defined by
\begin{equation}
\tilde{\Lambda}_{\ell} \equiv \Lambda_{\ell} \max_{i}|\tilde{I_i}| / N,
\end{equation}
which is less affected by the number of visibilities $N$ and the total flux density of the target source that should be close to the maximum value of the visibility amplitudes at Stokes $I$ (i.e.~$\max _i |\tilde{I}_i|$), following \citet{akiyama2016}.

The other additional term is total variation (TV) regularization with a regularization parameter $\Lambda _{t}$. A large $\Lambda_t$ leads to a sparse solution in the gradient domain -- a piecewise smooth solution. In this paper, we adopt the isotropic TV expression \citep[][]{rudin1992}, a typical form for two-dimensional images, defined by
\begin{eqnarray}
||{\bf x}||_{\rm tv}=\sum _i \sum _j \sqrt{|x_{i+1,j} - x_{i,j}|^2 + |x_{i,j+1} - x_{i,j}|^2}.
\end{eqnarray}
We have examined the effects of TV regularization on Stokes $I$ imaging in our previous work \citep[][]{ikeda2016,akiyama2016}, and TV regularization is also used in other applications, such as Doppler tomography \citep{uemura2015}. As with the $\ell _1$-norm, we use a normalized regularization parameter $\tilde{\Lambda}_{t}$ defined by
\begin{equation}
\tilde{\Lambda}_{t} \equiv 4 \Lambda_t \max_{i}|\tilde{I_i}|/N,
\end{equation}
again following \citet{akiyama2016}.

The Stokes $I$ image is solved with a non-negative condition (i.e. ${\bf I}\geq0$). The linear polarization image (henceforth $P$ image) is calculated from reconstructed $Q$ and $U$ images by $P=Q+iU$. In other words, Stokes $Q$ and $U$ images are solved independently. Since the Stokes $Q$ and $U$ images can be negative, we solve these images without the non-negative condition.

The optimization problem, equation (\ref{eq:opteq}), is convex, and therefore its solution is uniquely determined regardless of initial conditions. Many algorithms have been proposed to solve this problem. We adopt the fast iterative shrinking thresholding algorithm (FISTA), originally proposed by \citet{beck2009b} for $\ell _1$ regularization and by \citet{beck2009a} for TV regularization. We use a monotonic FISTA algorithm (MFISTA) designed for $\ell _1$+TV regularization (see Appendix \ref{sec:A} for details). 
%
%

\section{Imaging Simulations}\label{sec:3}
\subsection{Physically Motivated Models}\label{sec:3.1}
\begin{figure*}
	\centering
	\includegraphics[width=1.0\textwidth]{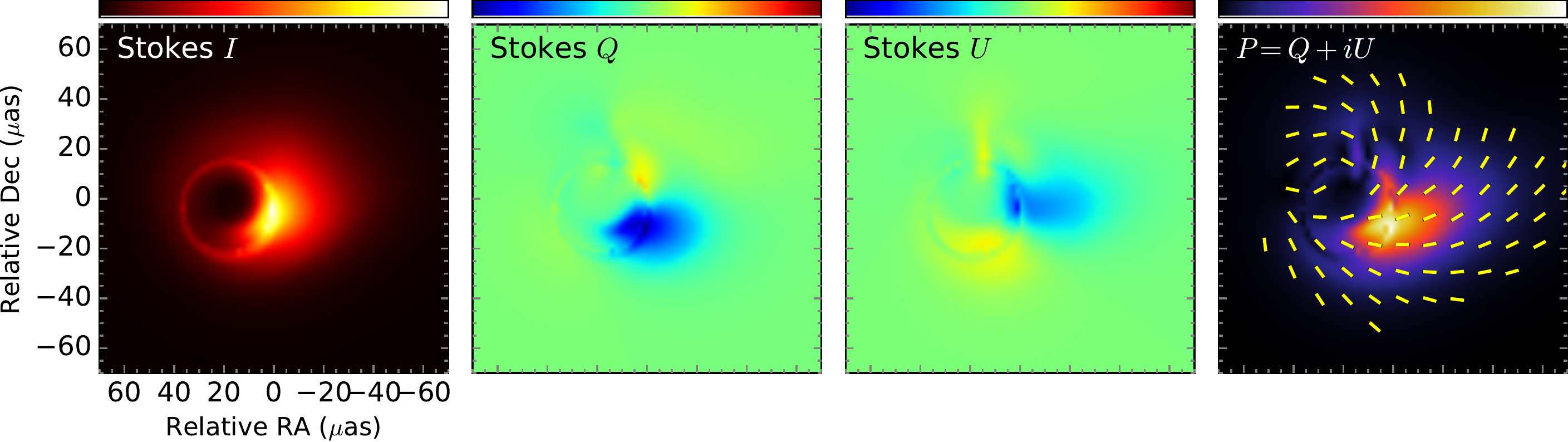}
	\includegraphics[width=1.0\textwidth]{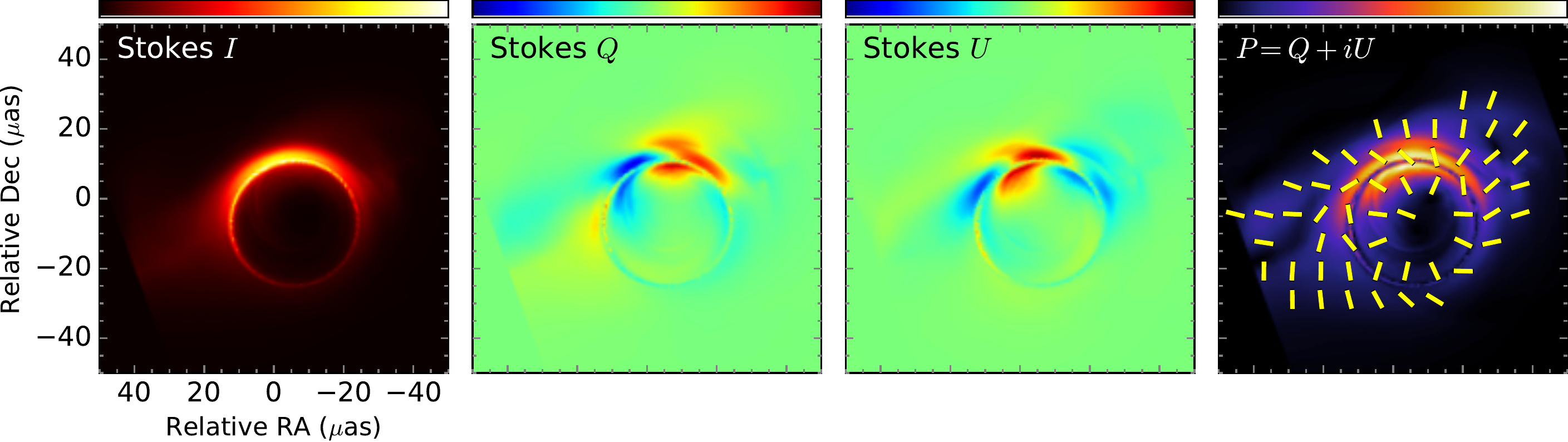}
	\caption{Two physical models of M87 for 1.3~mm emission on event-horizon scales, which are used for simulated observations. The upper panels show the forward jet model \citep{broderick2009,lu2014}, while the lower panels show the counter jet model \citep{dexter2012}. The leftmost panels show Stokes $I$ images with a linear color scale ranging from 0 to its peak intensity. The central two images are Stokes $Q$ and $U$ images with a linear color scale ranging from $-\max |P|$ to $\max |P|$, so that the center of the color scale (i.e. light green) shows an intensity of 0. The rightmost panels show $P$ images. The color contour indicates linear polarization intensity ($|P|$) with a linear scale from 0 to its peak, while the yellow bars show the EVPA distribution ($\arg (P)/2$).}
	\label{fig:modelimage}
\end{figure*}

In this paper, we adopt two physically motivated models of the 1.3~mm emission from M87 on event-horizon scales (Figure \ref{fig:modelimage}). In this paper, we focus on imaging the total intensity $I$ and linear polarization $Q$ and $U$ emission

The first model is a simple force-free jet model (hereafter, forward-jet model) in the magnetically dominated regime \citet{broderick2009,lu2014}. We adopt the model image presented in \cite{lu2014}, which is based on model parameters fitted to the spectral energy distribution of M87 and results of EHT observations at 1.3~mm \citep{doeleman2012}. The approaching jet is the dominant feature in this model. 

The second model (henceforth, counter-jet model) is based on results of GRMHD simulations \citep{dexter2012} and full polarimetric general relativistic radiation transfer calculations (\citealt{dexter2016}; Mo{\'s}cibrodzka, Dexter, Davelaar et al., in prep.). The dominant emission region is the counter jet illuminating the last photon orbit. 


\subsection{Simulated Observations}\label{sec:3.2}
\begin{figure}[t]
\centering
\includegraphics[width=1.0\columnwidth]{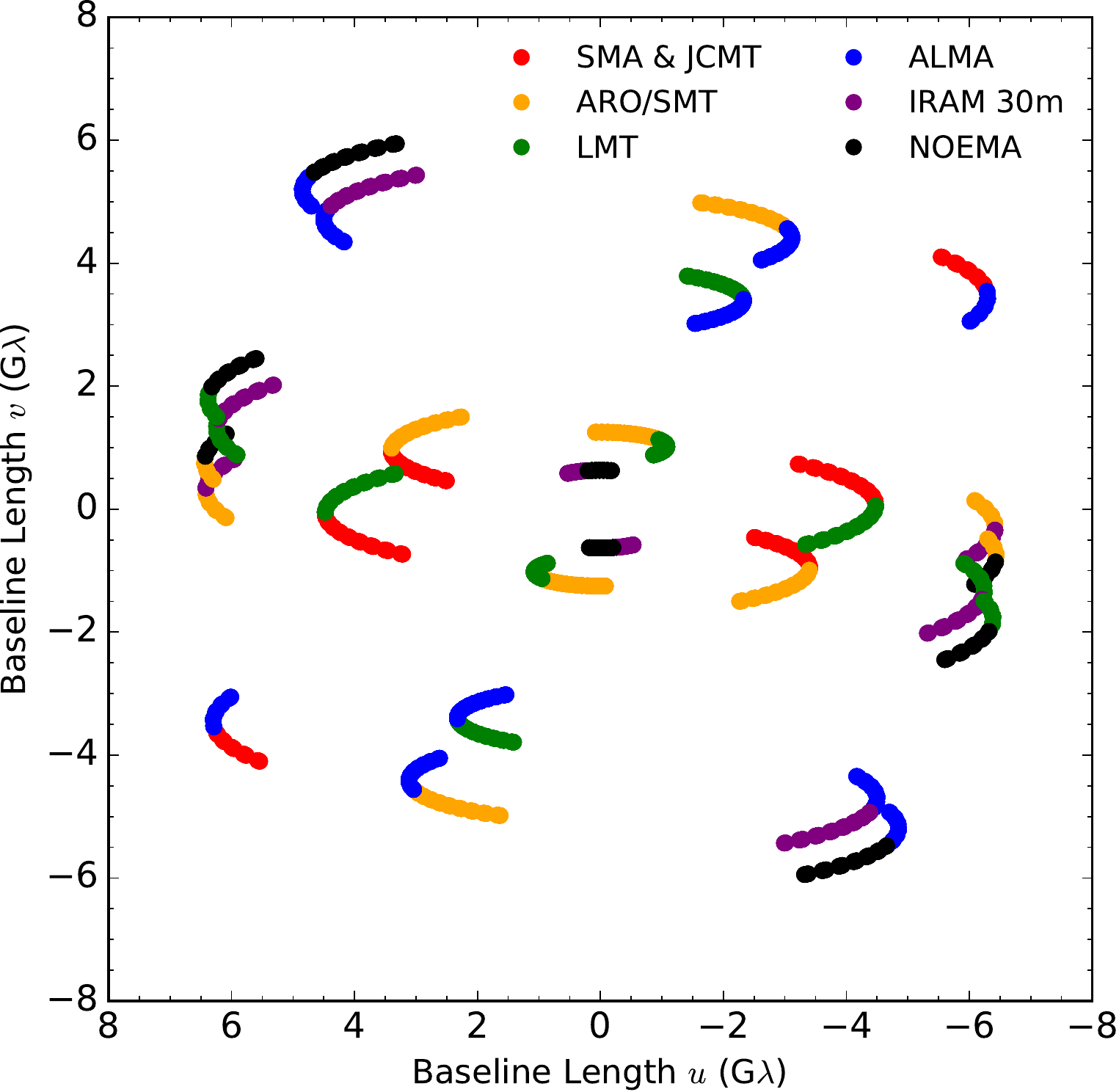}
\caption{The $uv$-coverage of the simulated observations. Each baseline is split into two colors to indicate the corresponding two stations.}
\label{fig:uv-coverage}
\end{figure}

We simulate observations of model M87 images with the EHT at 1.3~mm (230~GHz) using the MIT Array Performance Simulator (MAPS)\footnote{\url{http://www.haystack.mit.edu/ast/arrays/maps/}}. In most aspects, the data generation parameters are identical to those used in \citet{akiyama2016}, except that here we use an integration time of 10 sec. We simulate data for a six-station array with a band width of 3.5~GHz at each polarization, system temperatures described in the proposer's guid of 1-mm VLBI observations in ALMA Cycle 4, and a correlation efficiency of 0.7 that includes a quantization efficiency of 2-bit sampling and other potential losses. Observations are performed with an observational efficiency of 25\% in time, during a GST (Greenwich sidereal time) range of 13-0 hour. This GST range corresponds to a time range when M87 can be observed by either of two anchor stations of the EHT, the Atacama Large Millimeter/submillimeter Array (ALMA) or the Large Millimeter Telescope (LMT),  at an elevation greater than 20$^\circ$.  Here, we consider only thermal errors. See \citet{akiyama2016} for more details about the conditions of simulated observations. Fig.~\ref{fig:uv-coverage} shows the resultant $uv$-coverage of simulated observations. Note that the maximum baseline length of observations is 7.2~G$\lambda$, corresponding to $\lambda/D_{\rm max} =28.5$~$\rm \mu$as.

\subsection{Imaging}\label{sec:3.3}
We reconstruct Stokes $I$, $Q$ and $U$ images from simulated data-sets based on the method described in \S\ref{sec:2.2}. In addition, we attempt to solve images with the Cotton-Schwab CLEAN algorithm \citep[henceforth, CS-CLEAN;][]{schwab1984} using uniform weighting to evaluate the relative performance of our techniques in the exactly same way with \citet{akiyama2016}. We use an implementation of CS-CLEAN in the Common Astronomy Software Applications (CASA) package\footnote{\url{https://casa.nrao.edu/}}. We adopt a field of view (FOV) of 200~$\mu$as, gridded into 100~pixels in each of right ascension and declination for both models.  The resulting pixel size of $\sim2$~${\rm \mu}$as corresponds to a physical scale of $\sim 0.21$~$R_s$. 

The proposed method has two regularization parameters $\tilde{\Lambda}_{\ell}$ and $\tilde{\Lambda}_t$. Images  at each Stokes parameter were reconstructed at 5 regularization parameters for both $\tilde{\Lambda}_{\ell}$ and $\tilde{\Lambda}_t$, equally spaced in logarithmic steps in the range $10^{-1},...,10^{+3}$. In addition to employing $\ell _1$+TV regularization, we also explore pure $\ell _1$ and pure TV regularization separately (i.e., $\tilde{\Lambda}_t = 0$ or $\tilde{\Lambda}_{\ell} = 0$, respectively). 

We evaluate the goodness-of-fit for each image and then select the best-fit images with 10-fold Cross Validation \citep[henceforth CV;][]{akiyama2016}. The observational data (i.e. sampled visibilities) are randomly partitioned into 10 equal-sized subsamples. 9 of 10 subsamples are used in the image reconstruction as the {\it training set}, and we obtain the {\it trained} image. The remaining single subsample is used as the {\it validation set} for testing the model using $\chi^{2}$. The $\chi^{2}$ between the validation set and the image from the training set, which is so-called the {\it validation error}, is a good indicator of goodness-of-fit. For too small regularization parameters, the trained image would be over-fitted and too complicated, resulting in a large deviation between the trained image and the validation set (i.e. large validation error). On the other hand, for too large regularization parameters, the trained image would be too simple and not well-fitted to the training set, also resulting in a large validation error. Thus, reasonable parameters can be estimated by finding a parameter set that minimizes the validation error. We repeat the procedure by changing the subsample for validation data 10 times, until all subsamples are used for both training and validation. As a result, we get 10 validation errors. The validation errors are averaged and then used to determine optimal regularization parameters at each Stokes parameter that minimize the averaged validation error. Note that the final image is reconstructed by full sample of the observed visibilities.

To reduce the computational cost, we grid the observed visibilities with the classic cell-averaging method \citep[see][]{thompson2001} prior to imaging. We adopt a FOV size of 2~mas for gridding, corresponding to a grid size of $\sim 0.1$~G$\lambda$ in $uv$-space.

\subsection{Evaluation of the image fidelity}\label{sec:3.4}
We evaluate the quality of reconstructed images in two ways. First, we employ the normalized root mean square error (NRMSE) metric following \citet{chael2016} and \citet{akiyama2016}, defined as,
\begin{equation}
{\rm NRMSE}({\bf I},\,{\bf K}) = \sqrt{\frac{\sum_i |I_i-K_i|^2}{\sum_i |K_i|^2}},
\end{equation}
where ${\bf I}$ and ${\bf K}$ are the image to be evaluated and the reference image, respectively. For linear polarization images, we use the complex linear-polarization intensity (i.e. ${\bf P}={\bf Q}+i{\bf U}$) to evaluate NRMSEs. Since both model images have finer resolutions than is recoverable using the EHT, we adjust the pixel size of the reconstructed image to that of the model image with bi-cubic spline interpolation. Second, we measure structural dissimilarity \citep[]{wang2004} between the model and reconstructed images using the DSSIM metric adopted in previous work \citep{lu2014,fish2014}. Since both metrics show a similar trend, we show only the behavior of the NRMSE in the figures that follow.

Of potential interest for future EHT observations is to detect hypothesized blob-like flaring structure(s) in the accretion flow or jet due to partially heated or overdense plasma \citep[e.g.][]{broderick2006,doeleman2009a}. However, image reconstructions can generate artifacts that mimic bright components, making it difficult to identify such signatures accurately. A useful evaluation tool for imaging techniques is to identify how many bright blobs appear in the reconstructed image. The input model images do not contain flaring structures, so reconstructed images that show more than one cluster of pixels falsely recover blob-like features. We therefore also perform a cluster analysis on each image using Density-Based Spatial Clustering of Applications with Noise \citep[DBSCAN;][]{ester1996} to identify these false features. We configure DBSCAN as follows. The images contain a range of pixel brightness values; therefore, we cluster the pixels that have a brightness $>50\%$ of its peak intensity, of which separations are larger than 2 pixels $\rm \mu$as. 
Then we cluster the reconstructed images with the same parameters to find if false blob-like features (clusters) are obtained.

\section{Results}\label{sec:4}
\subsection{Stokes $I$ images}\label{sec:4.1}
\begin{figure*}
	\centering
	\includegraphics[width=1.0\textwidth]{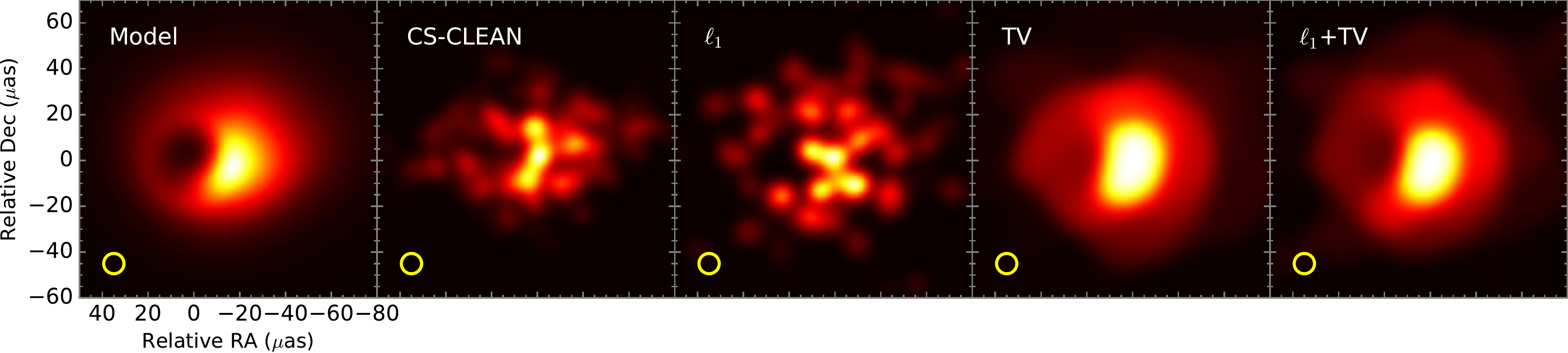}\\
	\includegraphics[width=1.0\textwidth]{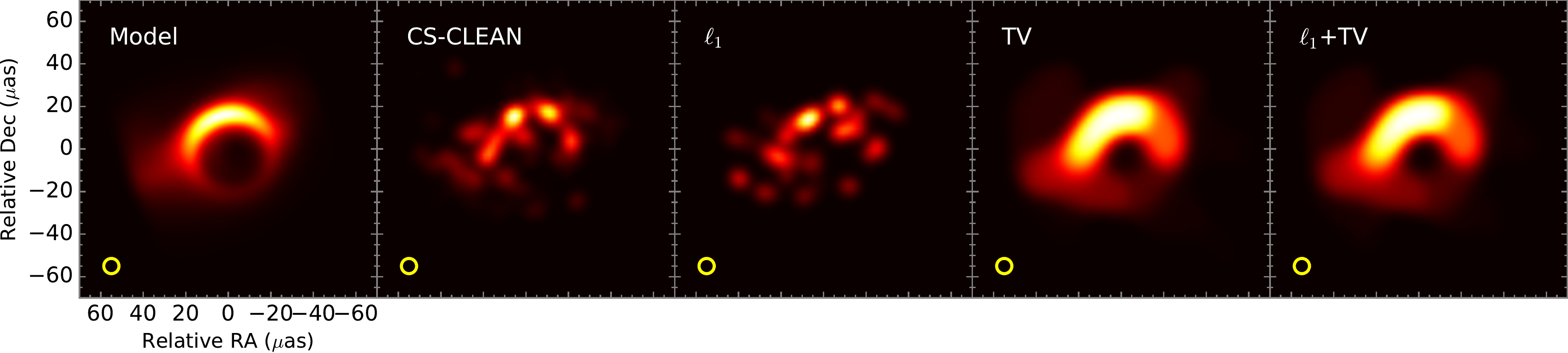}\\
	(a)
	\vspace{1em}
	\includegraphics[width=1.0\textwidth]{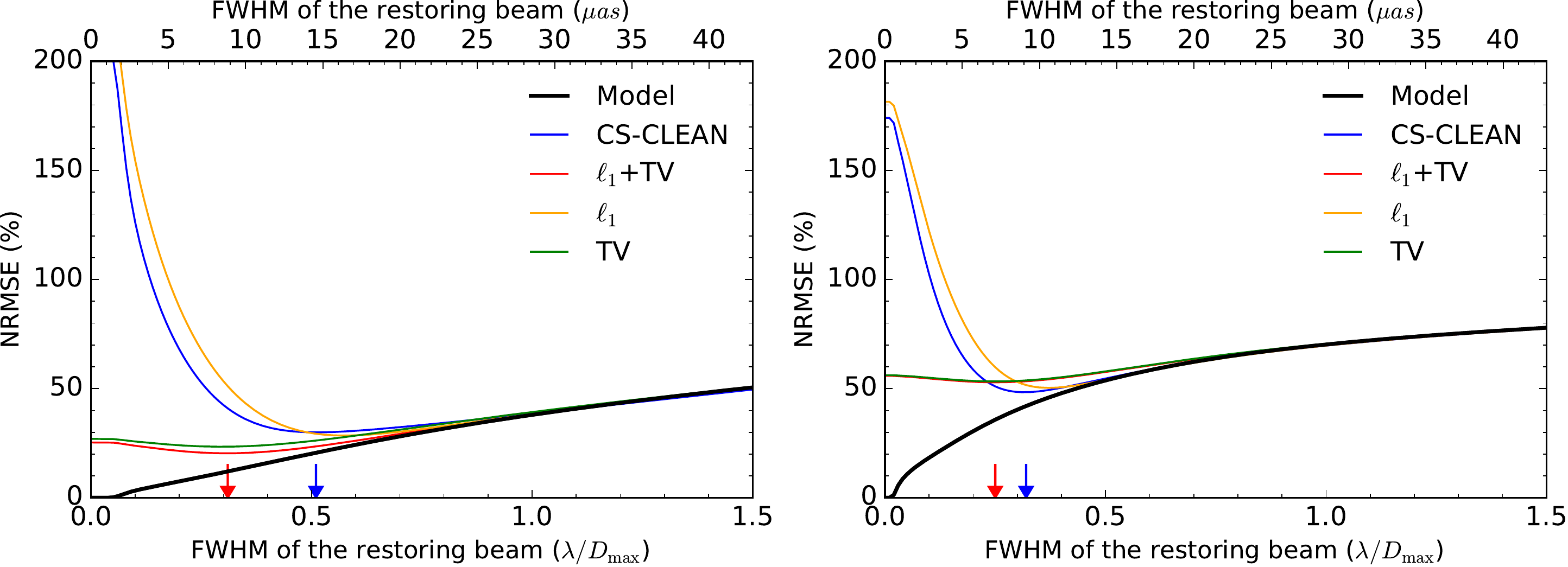}\\
	(b)
	\vspace{1em}
	\caption{The reconstructed Stokes $I$ images and evaluated metrics of the image fidelity for them. We adopt the contour scaling same to Figure \ref{fig:modelimage}. (a) The original model and reconstructed images for the forward-jet model (upper panels) and the counter-jet model (lower panels). All images are convolved with circular Gaussian beams with the FWMH sizes corresponding to diameters of the yellow circles, which coincide with the optimal resolutions for $\ell _1$+TV regularization shown in (b). (b) The NRMSE between the non-beam-convolved original model image and beam-convolved model/reconstructed images of the forward-jet (left) and counter-jet (right) models, as a function of the FWHM size of the convolving circular beam. The black curve indicates the NRMSE of the model image, while other curves indicate the NRMSEs of the reconstructed images. The red and blue arrows indicate the optimal resolution of $\ell _1$+TV regularization and CS-CLEAN, respectively, which minimize the NRMSE.}
	\label{fig:stokesi}
\end{figure*}

The results for Stokes $I$ images of $\ell _1$, TV and $\ell _1$+TV regularizations are shown along with the model and CS-CLEAN images in Figure \ref{fig:stokesi}.  We also plot the NRMSE metric for reconstructed images as well in the bottom panel, along with the full width at half maximum (FWHM) size of a convolving circular Gaussian beam. The best-case scenario --- the differences from the original input due solely to a loss of resolution, not to errors in reconstructing the image --- is shown by the black curve labeled ``Model'' following previous work \citep{chael2016,akiyama2016}. This is calculated by taking the NRMSE between the model image convolved with a circular Gaussian beam with a FWHM and the original (unconvolved) model image. The NRMSE of each of the reconstructed images, convolved with circular Gaussian beams, is shown in the bottom panel.


All techniques reconstruct Stokes $I$ images equally well on scales comparable to or greater than the diffraction limit. The NRMSEs of the reconstructed images only start to deviate from the model images in the super-resolution regime --- namely on scales finer than the diffraction limit. In this regime, the NRMSEs are different by techniques. $\ell _1$-regularization and CS-CLEAN have a common trend for both models. The minimum errors are achieved at a resolution of $\sim 30-50$\% of the diffraction limit, and then the NRMSEs show a rapid increase in errors at smaller scales, broadly consistent with previous studies on different model images and data products \citep{chael2016,akiyama2016}. This clearly shows that, on such small scales, the image is no longer sparse and breaks the underlying assumption of both techniques. In contrast, TV and $\ell _1$+TV regularizations show much more modest variations in the super-resolution regime. The minimum errors are achieved at a resolution of $\sim 25-30$\% of the diffraction limit, smaller than $\ell _1$-regularization and CS-CLEAN. In addition, the NRMSEs show only a slight increase in smaller scales. Both the TV and $\ell _1$+TV reconstructions produce images that have a smooth distribution similar to the model images, resulting in smaller errors than $\ell _1$-regularization and CS-CLEAN, even if the TV and $\ell _1$+TV are not convolved with a restoring beam.

A clustering analysis with DBSCAN shows that images with smoother regularizations ($\ell _1$+TV and TV) have only one cluster of bright imaging pixels regardless of resolution. The other two sparse techniques (CS-CLEAN and $\ell _1$) show more than one cluster in smaller resolutions, as clearly seen in Figure \ref{fig:stokesi} (a) for both models. Thus, even though all techniques have similar optimal resolutions and minimum NRMSEs for the counter-jet model, the bright emission has more than one clusters at optimal resolutions for CS-CLEAN and $\ell _1$. This indicates that sparse reconstructions with $\ell _1$-regularization and CS-CLEAN are more likely to misidentify flaring substructures. We also note that, simultaneously, this indicates that the NRMSE and DSSIM image fidelity metrics may not always be an appropriate indicator for goodness of feature reconstruction.


\subsection{Linear polarization images}\label{sec:4.2}
\begin{figure*}
	\centering
	\includegraphics[width=1.0\textwidth]{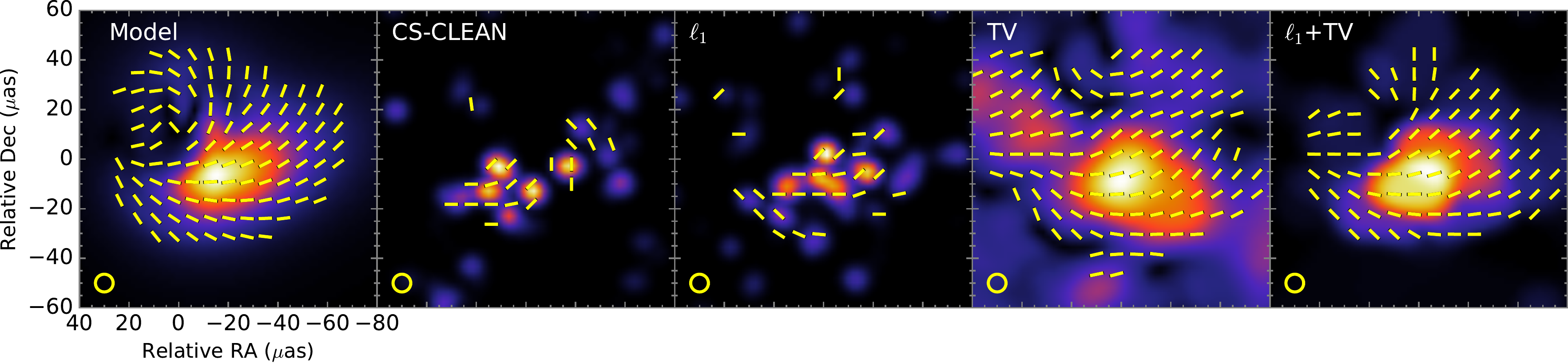}\\
	\includegraphics[width=1.0\textwidth]{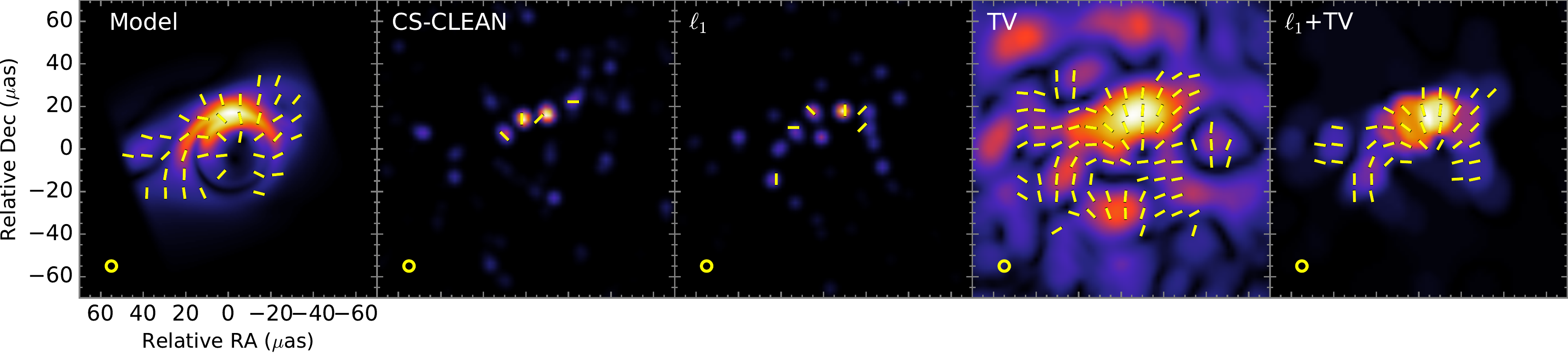}\\
	(a)\\
	\vspace{1em}
	\includegraphics[width=1.0\textwidth]{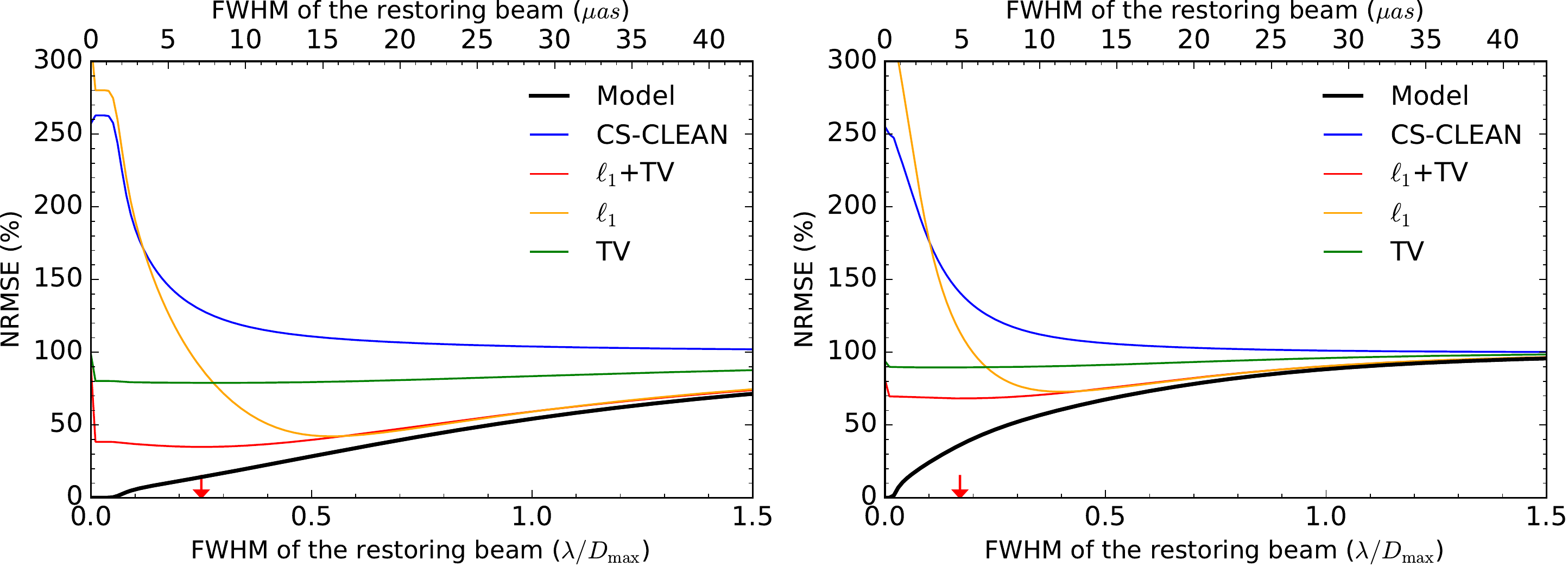}
	(b)\\
	\caption{The reconstructed linear polarization images (top) and evaluated metrics of the image fidelity for them (bottom).  Panels are as in Figure \ref{fig:stokesi}. Color contours and EVPAs of the top panel are shown as in Figure \ref{fig:modelimage}. The NRMSEs are calculated from the complex linear polarization intensity images ${\bf P}$, as described in \S\ref{sec:3.4}.} 
	\label{fig:lpol}
	\end{figure*}

The results for linear polarization images (Stokes $Q$ and $U$) are shown in Figure \ref{fig:lpol}. Similar to Stokes $I$ (Figure \ref{fig:stokesi}), we show the model/reconstructed images at the optimal resolution of $\ell _1$+TV regularization in panel (a) and the NRMSEs in panel (b). The NRMSEs show different trends for polarization than they do for Stokes $I$ \S\ref{sec:4.1} because Stokes $Q$ and $U$ can be negative and because the linearly polarized flux is significantly smaller than in total intensity, leading to a smaller signal-to-noise ratio in these simulations.

CS-CLEAN shows the highest NRMSE over almost the entire range of spatial scales, and its NRMSEs do not have a global minimum at resolutions smaller than 1.5$\lambda/D_{\rm max}$. Indeed, as shown in Figure 4, CS-CLEAN can recover only a tiny fraction of linearly polarized emission, and the reconstructed EVPA distribution is inaccurate. 

TV regularization, which shows good performances for Stokes $I$ imaging, is insufficient on its own for polarimetric imaging. The much lower signal-to-noise ratios in Stokes $Q$ and $U$ visibilities require a regularization parameter $\sim$10 times larger than for Stokes $I$ to minimize the validating error of 10-fold CV, resulting in image distributions that are much blurrier than the model images. The TV-regularized $P$ images reconstruct bright emissions better than CS-CLEAN, but there are a lot of artificial diffuse emissions that dominate the NRMSEs, as seen by the flat curves in Figure \ref{fig:lpol} (b).

The $\ell _1$ regularization exhibits better NRMSEs than CS-CLEAN for both models on most spatial scales. Since $\ell _1$ regularization suppresses the artificial diffuse emission seen in TV regularization, the achieved NRMSE is better than TV until at $\gtrsim 20$\% of the diffraction limit. However, as with Stokes $I$, the images become too sparse on scales smaller than $\sim 30-50$\% of the diffraction limit, causing a rapid rising in NRMSEs.

$\ell _1$+TV regularization provides reasonable linear polarization images with the most reasonable sparseness and smoothness, stably showing good performance across the entire range of spatial scales. The optimal resolution of $\sim 20-25$\% is comparable to Stokes $I$ and is the best among the four techniques. These results clearly shows that $\ell _1$+TV can achieve the best image fidelity among techniques presented in this paper not only for super-resolution imaging but also more general imaging on scales larger than diffraction limit.

\section{Discussions and Summary}
We have presented a new technique for full polarimetric imaging with radio interferometry using sparse modeling. As shown in \S\ref{sec:3} and \S\ref{sec:4}, $\ell _1$+TV regularization stably shows better performance than either $\ell _1$ or TV regularization alone, and than the most widely-used Cotton-Schwab CLEAN. This applies regardless of Stokes parameters. In particular, the superiority of the combined $\ell _1$+TV regularization is significant for linear polarizations on a wide range of spatial scales from super-resolution regimes to scales larger than the diffraction limit. 

Our technique can be applied to most existing radio interferometers whose data products are full complex visibilities in all four Stokes parameters. Although we did not image the circular polarization (Stokes $V$) in this work, our results suggest that $\ell _1$+TV regularization would likely achieve a better performance than the Cotton-Schwab CLEAN for circular polarization too, since it is mathematically similar to Stokes $Q$ and $U$ imaging.

We note future prospects for the application of our techniques to VLBI, including future EHT observations. In VLBI observations, the absolute phase information generally can not be obtained due to non-synchronized local oscillators and quite different atmospheric phase delays at different sites \citep[see][]{thompson2001}. In addition, at short-mm/sub-mm wavelengths, even the source visibility phase cannot be measured due to the rapidly varying atmospheric delays. In VLBI, the visibility phase is traditionally calibrated based on phase closure, using the self-calibration technique with hybrid/differential mapping \citep[e.g.][]{walker1995}. The systematic phase errors derived using Stokes $I$ data can be applied to Stokes $Q$ and $U$, since station-based systematic errors should be the same among the Stokes parameters. Our techniques can be applied to the VLBI data after self-calibrating Stokes $I$ data with the traditional hybrid/differential mapping or the Stokes $I$ image obtained with new state-of-art imaging techniques based on closure quantities such as Maximum Entropy Methods \citep[e.g.][]{buscher1994,chael2016}, a patch prior \citep[CHIRP;][]{bouman2015}, sparse modeling \citep{akiyama2016} and PRECL \citep{ikeda2016}, which have been developed for optical interferometers and/or the EHT. In a forthcoming paper, we will evaluate the performance of the proposed technique for data with station-based systematic phase errors, which are common in VLBI.

We also note that there is a new method for Stokes I and linear polarization, very recently proposed in \citet{chael2016}, which is designed for VLBI. This method solves first for the Stokes-$I$ images from visibility amplitudes and closure phases at Stokes $I$. Then, the linear polarization images are solved using the reconstructed Stokes-$I$ images and complex polarimetric ratios (ratios of the Stokes $Q$, $U$ visibilities to the Stokes $I$ visibilities).  All of these VLBI observables are robust against station-based phase errors. Through the above processes, Stokes $I$ visibility phases are recovered from imaging with visibility amplitudes and closure phases, and the visibility phases at Stokes $Q$ and $U$ are phase-referenced from Stokes $I$ through the polarimetric ratio. The Stokes-$I$ and polarization intensity (i.e. $|P|$) images are regularized by the entropy term of MEM, while the EVPA distribution (i.e. $\arg (P)/2$) is regularized independently by a smooth regularization term such as TV. \citet{chael2016} demonstrate that this method can also achieve a better fidelity and superior optimal resolution than the Cotton-Schwab CLEAN. An advantage of this technique is that it can simultaneously reconstruct $I$, $Q$ and $U$ images from robust VLBI observables. In addition, the reconstructed images strictly satisfy $|I|>|P|$, which can suppress artifacts in $P$ images in regions where $I$ is not bright. The disadvantage of this technique is that the optimization problem is highly non-linear and non-convex, and that the solution can potentially be initial-condition dependent not only in Stokes $I$ but also in Stokes $Q$ and $U$. Furthermore, $Q$ and $U$ images are reconstructed from the polarimetric ratio that can have larger uncertainties than the Stokes visibilities, particularly at long baselines, limiting the dynamic range, image sensitivity and optimal spatial resolution. An alternative, mathematically equivalent way --- phase-referencing with self-calibration techniques --- will avoid such disadvantages in polarimetric imaging. 

Future work for the techniques proposed in this paper will include other sparse regularizations for multi-resolution imaging, such as $\ell _1$+wavelet/curvelet transformation  \citep[e.g.][]{li2011,carrillo2014,garsden2015,dabbech2015}. 
In addition, the application of, and experimentation with other forms of TV would be important. We have been using the most widely-used isotropic TV  \citep[][]{rudin1992} for TV regularization, which preserves sharp edges in the image. This would be useful for optically thick objects like stars, but might not be optimal for emission from optically thin objects that have smoother edges in general. An alternative form of sparse regularization in the gradient domain that favors smoother edges is given by, for instance,
\begin{eqnarray}
||{\bf x}||_{\rm tv^2}=\sum _i \sum _j \left(|x_{i+1,j} - x_{i,j}|^2 + |x_{i,j+1} - x_{i,j}|^2 \right).
\end{eqnarray}
This function is also convex, similar to the TV term adopted in this work, and can be an alternative option. The performance of these sparse regularizers has not yet been fully evaluated for super-resolution imaging of compact objects with complicated structures on scales comparable to the diffraction limit, as is expected for black hole shadow imaging with the EHT. We will study this issue both for imaging with the full-complex visibility and closure quantities as an extension of this work and our previous work \citep{honma2014,ikeda2016,akiyama2016}.

A relevant issue of our proposed methods is the computational time, most of which is spent in determining the optimum parameters for the regularization terms. Since we adopt 10-fold CV for determining regularization parameters, we need to do image reconstruction 11 times for each set of regularization parameters. This is not serious for imaging simulated data sets in this paper, which takes about a few hours in total for each Stokes parameter, although it would be a relevant issue for imaging larger data sets or wider field-of-views. Recently, \citet{obuchi2016a} have proposed an accurate approximation of the validating error for $n$-fold CV for LASSO, which can be derived from the image reconstruction of full data sets. A similar approximation for TV regularization has been also derived very recently \citep{obuchi2016b}. These approximations may allow validating errors to be estimated by imaging the full data set just once at each set of regularization parameters. We will implement these estimators for our algorithm, which will significantly reduce the whole computational time ($\sim$ 10 times shorter for 10-fold CV). We will also work on optimizing and accelerating the MFISTA algorithms by parallel computing such as GPGPU (General-Purpose computing on Graphics Processing Units). This will be helpful for extending our works to wider-FOV imaging or imaging of much larger data sets with many more stations than VLBI networks, such as ALMA.

\acknowledgements
K.A. thanks Dr.~Michael~D.~Johnson, Dr.~Lindy~Blackburn, Katherine~L.~Bouman and Andrew~Chael for many fruitful discussions and constructive suggestions on this work. K.A. and this work are financially supported by the program of Postdoctoral Fellowships for Research Abroad at the Japan Society for the Promotion of Science (JSPS). M.P. acknowledges support from the NASA Massachusetts Space Grant Consortium and the National Science Foundation's (NSF) Research Experiences for Undergraduates program. M.M. acknowledges support from the ERC Synergy Grant (Grant 610058). Event Horizon Telescope work at MIT Haystack Observatory and the Harvard-Smithsonian Center for Astrophysics is supported by grants from the NSF (AST-1440254, AST-1614868) and through an award from the Gordon and Betty Moore Foundation (GMBF-3561). Work on sparse modeling and Event Horizon Telescope at the Mizusawa VLBI Observatory is financially supported by the MEXT/JSPS KAKENHI Grant Numbers 24540242, 25120007 and 25120008. 

\appendix

\section{Monotonic FISTA algorithm}\label{sec:A}

We show the details of the algorithms which were used to
solve equation (\ref{eq:opteq}). Our algorithms are variations
of the monotonic FISTA (MFISTA) which was introduced in
\citet{beck2009a,beck2009b}. We first show the general framework
of MFISTA and discuss how we applied it for our problem.

\subsection{General framework of MFISTA}

The general form of the problem is defined as follows,
\begin{equation}
\label{eq:define F}
\min_{\V{x}\in C}\{F(\V{x})\equiv f(\V{x}) + g(\V{x})\},
\end{equation}
where, $\V{x}\in\Re^n$ and $C\subseteq\Re^n$ is some closed
subset of $\Re^n$. The properties assumed for $f(\V{x})$ and $g(\V{x})$ are summarized below.
\begin{itemize}
	\item $f(\V{x}):\Re^n\rightarrow \Re$, is a convex function of $\V{x}$. It is continuously 
	differentiable, and the gradient $\nabla f(\V{x})$ is Lipschitz continuous, where $L(f)$ denotes the Lipschitz constant of $\nabla f(\V{x})$ .
	\item $g(\V{x}):\Re^n\rightarrow (-\infty,\infty]$, is
	a convex function of $\V{x}$. It is not necessarily smooth.
\end{itemize}

The pseudo code of MFISTA is summarized in algorithm \ref{algo:general MFISTA}.

\begin{algorithm}[H]
	\caption{MFISTA}
	\label{algo:general MFISTA}
	\begin{algorithmic}[1]
		\State Take $\V{x}_0\in\Re^{n}$, $L_0\in\Re$, and $\eta > 1$.
		\State $\bm{y}_1 \leftarrow \V{x}_0$, $t_1 \leftarrow 1$.
		\For{$k \geq 1$}
		\State {$L_k \leftarrow \mbox{\sc Initialize}(L_{k-1};\V{y}_k)$} \Comment{See \ref{algo:initialize}}
		\State $\V{z}_k \leftarrow p_C(\V{y}_k;L_k)$
		\Comment{$p_C(\V{y}_k;L_k)$ is defined in equation (\ref{eq:proximal}) }
		\State $t_{k+1}\leftarrow\frac{1+\sqrt{1+4t_k^2}}{2}$
		\If{$F(\V{x}_k)>F(\V{z}_k)$}
		\State $\V{x}_k \leftarrow  \V{z}_k$
		\State $\V{y}_{k+1}\leftarrow \V{x}_k + \frac{t_{k}-1}{t_{k+1}}(\V{x}_k - \V{x}_{k-1})$
		\Else
		\State $\V{x}_k \leftarrow  \V{x}_{k-1}$
		\State $\V{y}_{k+1}\leftarrow \V{x}_k + \frac{t_{k}}{t_{k+1}}(\V{z}_k - \V{x}_k)$
		\EndIf
		\If {converged}
		\State {\bf break}
		\EndIf
		\EndFor
	\end{algorithmic}
\end{algorithm}

If the upper bound of the Lipschitz constant $L(f)$ is known, $L$ is set to the upper
bound and ${\mbox{\sc Initialize}}(L_{k-1},\V{y}_k)$ can be omitted. Otherwise we need to find an 
appropriate value of $L_k$.
Let us define a function $Q(\bm{x},\bm{y};L)$ as follows
\begin{equation}
Q(\bm{x},\bm{y};L) = f(\V{y}) + 
\bigl\langle 
\V{x}-\V{y},
\nabla f(\V{y})
\bigr\rangle
+\frac{L}{2}
\bigl\|
\V{x} -\V{y}
\bigr\|_2^2
+g(\V{x}),
\end{equation}
where $\langle \cdot, \cdot \rangle$ denotes the inner product.
The function $p_C(\V{y};L)$ is the proximal map which is defined as follows,
\begin{equation}
\label{eq:proximal}
p_C(\V{y};L) = \argmin_{\V{x}\in C}
Q(\bm{x},\bm{y};L) 
=
\argmin_{\V{x}\in C}
\Biggl\{
\frac{L}{2}
\Bigl\| \V{x} - 
\bigl(
\V{y} - \frac{1}{L}\nabla f(\V{y})
\bigr)
\Bigr\|_2^2 + g(\V{x})
\Biggr\}.
\end{equation}
The practical form of the proximal map depends on the definition of $g(\V{x})$.

The procedure $\mbox{\sc Initialize}(L;\V{x})$ is defined as follows using  
$Q(\bm{x},\bm{y};L)$ and $p_C(\V{y};L)$.
\begin{algorithm}[H]
	\caption{Initial $L$}
	\label{algo:initialize}
	\begin{algorithmic}[1]
		\Procedure{Initialize}{$L;\V{x}$}
		\Repeat
		\State{ $L \leftarrow \eta L$}
		\Until
		{$\displaystyle F( p_C(\V{x}; L) )
			\le
			Q(p_C(\V{x}; L),\V{x}; L)$}
		\State{{\bf return} $L$}
		\EndProcedure
	\end{algorithmic}
\end{algorithm}

MFISTA only uses the gradient of $f(\V{x})$ and is known to have a fast convergence rate. 
Let $\V{x}^\ast$ be the optimal point of the problem in equation (\ref{eq:opteq}). The
MFISTA algorithm has the following property \citep{beck2009a,beck2009b},
\begin{equation}
F(\V{x}_k)-F(\V{x}^\ast)
\leq
\frac{2\alpha L(f)\|\V{x}_0 - \V{x}^\ast\|_2^2}{(k+1)^2}, 
\hspace{2em}
\forall \V{x}^\ast\in C.
\end{equation}

\subsection{Applying MFISTA for Polarimetric Imaging}

We explain how we applied MFISTA to solve equation (\ref{eq:opteq}). 

For the Stokes $I$ image,  ${\bf S}={\bf I}$ and $I_i \ge 0$. Since $\|{\bf I}\|_1 = \sum_i I_i$ 
holds, we can apply MFISTA by defining $f(\cdot)$, $g(\cdot)$ and $C$ as follows,
\begin{equation}
f({\bf I}) = \| \tilde{\bf I} - {\bf FI}\|_2^2 + \Lambda_{\ell} 
\sum_i I_i,
\hspace{2em}
g({\bf I}) = \Lambda_t \|{\bf I}\|_{\rm tv},
\hspace{2em}
C = \{{\bf I} ~|~ I_i\ge 0, \, \mbox{for }\forall i\}.
\end{equation}
The form of the proximal map $p_C(\V{y};L)$ for the case $g(\V{x}) = \|\V{x}\|_{\rm tv}$ has 
been discussed in \citet{beck2009a}. We used their Fast Projected Gradient (FGP) method
restricting $I_i\ge 0$.

For the Stokes $Q$, $U$ and $V$ image, each component can take negative value. 
MFISTA can be applied by defining $f(\cdot)$ and $g(\cdot)$ as follows,
\begin{equation}
f({\bf S}) = \| \tilde{\bf S} - {\bf FS}\|_2^2,
\hspace{2em}
g({\bf S}) = \Lambda_1\|{\bf S}\|_1 + \Lambda_t \|{\bf S}\|_{\rm tv},
\hspace{2em}
{\bf S}= {\bf Q},\,~{\bf U}~\mbox{or}~{\bf V}.
\end{equation}
The proximal map $p_C(\V{y};L)$ for this case can also be realized by a slight modification of FGP.


\end{document}